\documentclass[12pt]{article}
\usepackage{scicite,times,amsmath,amssymb}
\usepackage{epsfig,graphicx}
\newcommand{\smfrac}[2]{\mbox{$\frac{#1}{#2}$}}
\def\>{\rangle}
\def\<{\langle}
\newcommand{\eq}[1]{Eq.~(\ref{eq:#1})}
\def\e{\epsilon}

\def\untrust{\rho_{\rm 0}}

\def\whpgood{\tilde{\gamma}}

\def\duzomniejsze{<\kern-.7mm<}
\def\duzowieksze{>\kern-.7mm>}

\def\textbf#1{{\bf #1}}

\def\bep{\begin{proposition}}
\def\eep{\end{proposition}}

\def\beq{\begin{equation}}
\def\eeq{\end{equation}}
\def\be{\begin{equation}}
\def\ee{\end{equation}}
\def\bea{\begin{eqnarray}}
\def\eea{\end{eqnarray}}
\def\beqa{\begin{eqnarray}}
\def\eeqa{\end{eqnarray}}
\def\ben{\begin{eqnarray}}
\def\een{\end{eqnarray}}

\newcommand{\bei}{\begin{itemize}}
\newcommand{\eei}{\end{itemize}}
\newcommand{\bee}{\begin{enumerate}}
\newcommand{\eee}{\end{enumerate}}

\def\hcal{{\cal H}}
\def\pcal{{\cal P}}
\def\acal{{\cal A}}

\def\mcal{{\cal M}}
\def\zcal{{\cal Z}}

\def\tr{{\rm Tr}}

\def\>{\rangle}
\def\<{\langle}

\def\ot{\otimes}

\def\ot{\otimes}  

\def\ep{\epsilon}

\def\ap_nr{$Sym(\hcal^{\ot n},|\theta\>^{\ot n-r})$}
\def\mix_ap_nr{$\pcal(Sym(\hcal^{\ot n},|\theta\>^{\ot n-r}))$}

\newtheorem{lemma}{Lemma}

\newtheorem{proposition}{Proposition}
\newtheorem{theorem}{Theorem}
\newtheorem{definition}{Definition}

\newtheorem{corollary}{Corollary}

\newtheorem{fact}{Fact}

\def\Sym{{\rm Sym}}
\def\Li{\langle L_{i} \rangle_{\sigma}}
\def\Romr{\varrho^{(\sigma)}_{2m,r}}
\def\Liemp{\langle L_{i} \rangle_{emp}}
\def\Sgm{\langle\Sigma\rangle_{\sigma}}  
\def\SgmAemp{\langle\Sigma\rangle_{emp}^{(m)}}
\def\SgmAAemp{\langle\Sigma\rangle_{emp}^{(m+n)}} 
\def\SgmBemp{\langle\Sigma^{ind}\rangle_{emp}^{(m)}} 

\def\SgmAAempnon{\langle\Sigma\rangle_{emp}} 
\def\SgmBempnon{\langle\Sigma^{ind}\rangle_{emp}} 

\def\lcal{{\cal L}} 
\def\vcal{{\cal V}}
\def\bec{\begin{corollary}}
\def\eec{\end{corollary}}

\def\bel{\begin{lemma}}
\def\eel{\end{lemma}}
\def\bet{\begin{theorem}}
\def\eet{\end{theorem}}

\def\almpower{\rho^{(\sigma)}_{n,r}}

\newenvironment{sciabstract}{%
\begin{quote} \bf}
{\end{quote}}

\newcounter{lastnote}

\title{Unconditional privacy over channels which cannot convey quantum information}
\author{Karol Horodecki$^{(1)}$, Micha\l{} Horodecki$^{(1)}$, Pawe\l{} Horodecki$^{(2)}$,
Debbie Leung$^{(3)}$, \\and Jonathan Oppenheim$^{(4)}$
\\
\normalsize{$^{(1)}$Department of Math, Physics and Computer
Science, University of Gda\'nsk, 80--952 Gda\'nsk, Poland,}\\
\normalsize{$^{(2)}$Faculty of Applied Physics and Math,
Technical University of Gda\'nsk, 80--952 Gda\'nsk, Poland}\\
\normalsize{$^{(3)}$Institute for Quantum Computing, University of 
Waterloo, Waterloo, Ontario, N2L1N8, Canada}\\
\normalsize{$^{(4)}$Deptartment of Applied Mathematics and
Theoretical Physics, University of Cambridge U.K.}\\
\\
}

\begin{document}
\baselineskip24pt


\maketitle 
\begin{sciabstract}
By sending systems in specially prepared quantum states, two parties can communicate 
without an eavesdropper being able to listen.  The technique, called quantum 
cryptography, enables one to verify that the state of the quantum system has not 
been tampered with, and thus one can obtain privacy regardless of the power of the eavesdropper. 
All previous protocols relied on the ability to faithfully send quantum states.  
In fact, until recently, they could all be reduced to a single protocol where 
security is ensured though sharing maximally entangled states.  Here we show 
this need not be the case -- one can obtain verifiable privacy even through 
some channels which cannot be used to reliably send quantum states.

\end{sciabstract}

The nature of quantum systems and our ability to manipulate the state they are in has had a radical
impact on the field of information theory and computation. A quantum computer can solve problems which
a classical computer cannot, and photons prepared in special states can be used to 
obtain privacy between two individuals sharing a fiber-optic channel -- a situation impossible 
classically.  
Researchers in quantum information theory are trying to understand  what 
aspects of quantum states and manipulations are responsible for the power of 
quantum computation and cryptography.  

In the case of cryptography the ability to faithfully send arbitrary 
quantum states \cite{BDSW1996} appeared to lay at
the heart of obtaining privacy.  In the original protocol, BB84~\cite{bb84}, two-level quantum systems such as photons
were faithfully sent in eigenstates of one of two complementary basis, which allows both privacy and the 
faithful sending of quantum states.  Equivalently, entanglement based schemes~\cite{Ekert91} relied on the faithful
distribution of maximally entangled pure states, which again allows the transmission of arbitrary states. 
In reality, the quantum states used in such protocols are so fragile that interaction with the 
environment (or the eavesdropper) causes them to rapidly 
decohere. However, if there is not too much noise, 
one can perform quantum error correction on the states, as one does in quantum computation, 
or post-processing on the raw key, to overcome the noise.  
The environment and the eavesdropper then become decoupled from the quantum states and 
the two parties can then
obtain privacy. 

Since all known protocols achieve privacy by decoupling the eavesdropper from the sent states,
there was much reason to assume that this is necessary.
This implied that the faithful sending of arbitrary quantum states (such as halves of maximally entangled states)
appeared to be a necessary precondition for privacy. 
In other words, all previous cryptographic schemes are qualitatively equivalent to each other,
and equivalent to distilling pure state entanglement. 
The first step in showing that this need not be the case
was in \cite{HHHO03} in the scenario where trusted states are given to the parties.
There, we obtained the most general state which can produce a private 
key upon measurement.  One can then recast all of quantum cryptography as a protocol which distills
these {\it private states} under local operations and classical communication (LOCC).  It was then shown
that there exist private states which are not equivalent to pure state entanglement.  In fact,
they can be produced from channels which have zero capacity~\cite{HHH99,UPBII1999} -- the channels 
cannot be used to faithfully send arbitrary quantum states, but they can produce
states which are private.  However, a key ingredient remained.  For quantum key distribution (QKD)
it is not enough for two parties to share a private state, they must be able to verify this privacy.  
One imagines a scenario where the eavesdropper actually gives the two parties the states, or the
parties produce the states through a channel which the eavesdropper can tamper with.  One must
be able to verify that one indeed holds a private state and not something else.

Here, we provide a protocol which allows two parties (Alice and Bob) to verify that they indeed possess
private states using only LOCC.  This works for all private states, even those which can
be created from zero-capacity channels, thus allowing us
to obtain security over channels which cannot be used to send quantum information.  The protocol 
is thus inequivalent to the original schemes.  We 
previously~\cite{HLLO05} had introduced a protocol which worked over channels which could have
arbitrary small capacity, but the protocol cannot be extended to the case where the capacity is strictly
zero.
Here, we will simply sketch the proof of security of our protocol. 
The technical details are contained in the appendix as well as in \cite{HHHLO06}.

Let us recall that there are two scenarios for QKD.  In entanglement based schemes, an adversary gives
states to Alice and Bob and they distill pure entanglement in the form of the maximally 
entangled state
$|\Phi_d\> := \smfrac{1}{\sqrt{d}} \sum_{i=1}^d |i\>_A |i\>_B$ where
$\{|i\>\}$ is a computational basis for the local systems $A$ and $B$
possessed by Alice and Bob respectively.  They then verify that they indeed possess states very close to
this form, and then measure
in the computation basis to produce a secure key.  One also has prepare and measure protocols, where
Alice prepares a quantum state, sends it to Bob who then measures it in some basis.  They then examine
the results to verify that the sent states were not overly tampered with, and then 
perform classical post-processing 
on the results to obtain a key.  
The two schemes are equivalent in the sense that current prepare and measure schemes can be reduced to
protocols which rely on the distillation and verification of maximally entangled states as 
shown in \cite{Shor-Preskill}.
In \cite{HHHO03} it was shown that one could consider more general schemes where one
considered protocols which rely on the distillation of states of the form
%
%
\bea
        \gamma_d^U & = & U (|\Phi_{dAB}\>\<\Phi_{dAB}| \ot \rho_{A'B'}) U^\dagger 
\label{eq:gamma}
\\
        U & = & \sum_{ij} |ij\>\<ij|_{AB} \ot U_{ijA'B'}
\label{eq:twist}
\eea
and viewing any protocol as the distillation and verification of such private states.  Here $\rho_{A'B'}$
is an arbitrary ancilla,  the $U_{ijA'B'}$ are arbitrary unitaries on it, and $U$ is called
{\it twisting}.

\begin{figure}[tbp]
\label{fig}
\begin{center}
\epsfig{figure=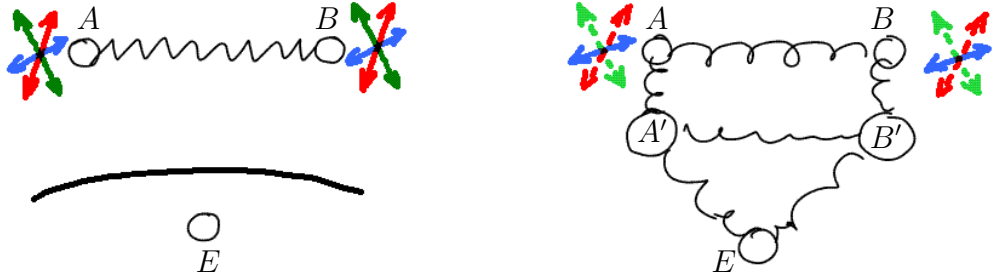,width=0.9\textwidth}
\put(-360,95){\(A\)}
\put(-270,95){\(B\)}
\put(-145,95){\(A\)}
\put(-60,95){\(B\)}
\put(-148,52){\(A'\)}
\put(-60,50){\(B'\)}
\put(-315,3){\(E\)}
\put(-120,3){\(E\)}
\caption{For a channel with non-zero capacity (left hand side),
maximally entangled states can be distributed such that 
measuring in any direction will produce a private key.   
The eavesdropper is completely decoupled from the state.
For the channel on the right hand side with zero capacity,
the distributed states only produce a private key if the 
measurement is made along the blue axis. It can be shown that the eavesdropper
must know at least one bit of information, but due to the ancilla on 
$A'B'$ (the ``shield''), 
the bit of information is not about the key.
} 
\end{center}
\end{figure}

We now give the protocol for verifying private states, and prove its security.  
The protocol is a twisted version
of verification schemes of $|\Phi_d\>$, and in the spirit of \cite{Shor-Preskill} 
we will prove security of our protocol by reducing
it to security of the protocol due to Lo and Chau~\cite{LC99}.
Let us recall that the Lo-Chau protocol is as follows:

(1) Alice can locally prepare $n$ systems in the state $|\Phi_2\>$ and distribute Bob's share to him
through an untrusted channel where 
the eavesdropper can attack all of Bob's share at once before it gets to him. After this step, they share 
the state $\untrust$.

(2)  Perform tests (via public but authenticated discussion) on $\untrust$
by randomly selecting $m_x$ and $m_z$ systems, and measuring  $\sigma_z
\otimes \sigma_z$ to estimate the bit error rate $\e_x$, and  $\sigma_x
\otimes \sigma_x$ to measure the phase error rate $\e_z$.  Here, the $\sigma$ are the
standard Pauli matrices.  The error rates essentially tell us how far $\untrust$ deviates
from a maximally entangled state. 

(3) Based on the results of the test, the parties perform an appropriate entanglement purification protocol (EPP) to 
$\untrust$ and output a state $\whpgood$ which will be close to the maximally entangled state with high probability.  
One doesn't need to know
the exact form of $\untrust$, but only the error rates.
 
(4) Generate a key by measuring $\whpgood$ locally. 
    The key can have varying size (depends on the error rate), and zero
    key-length means ``abort QKD.''

%

The security of this protocol rests on the fact that the estimates $\e_x,\e_z$ 
of the two error rates by 
random sampling will converge with high probability to their expectation values over the entire 
initial state $\untrust$
thus ensuring that the final 
state $\whpgood$ is close to maximally entangled. For small $\delta$ and $m_z <
(\smfrac{2\delta^2}{1{+}2\delta^2}) n$, we have for example \cite{LC99}
\bea
    {\rm Pr}( \, |\e_{zP}{-}\e_{z}| \geq \delta) \leq 2  e^{- 2 m_z \delta^2 }
\label{eq:lcrate}
\eea
where $\e_{zP}$ is the expectation value of the phase error rate.
This result is from sampling theory and can be found as Proposition 1
in
the appendix.  
%

We now wish to modify this protocol so that we can use it to verify private states, 
which for the moment we take to be many copies of $\gamma_2^U$.  In 
\cite{HHHO03,huge-key,HPHH05} examples of such states were given which result from
zero-capacity channels (i.e. they are {\it bound entangled}\cite{bound}), and
thus our protocol will work over such channels.

Since private states are twisted maximally entangled states, we could achieve verifiable privacy,
by {\it untwisting} the private state before each step of the protocol, 
so that we are just acting the above protocol on the maximally entangled state.
We would thus need to modify the protocol as follows: 

(2') Apply untwisting $U^{\ot n \dagger}$ to $\untrust$, then estimate
$\e_x$ and $\e_z$ on the $(AB)^{\ot n}$ systems as in the original step (2), 
and finally reapply $U^{\ot n}$.

(3') Apply untwisting $U^{\ot n \dagger}$, measure out a ``raw-key''
in the computational basis of the remaining $n{-}m_x{-}m_z$ systems.

(4') Perform error correction and privacy amplification on the
raw-key via public discussion.

Such a protocol is unfeasible since $U$ may be a global unitary and cannot be done using 
only LOCC.  However, it is secure, since if we were able to perform the twisting and
untwisting, the only difference between this protocol and that of the Lo-Chau one
is that classical privacy amplification~\cite{BBCM95} and error correction is used instead of 
the entanglement purification protocol (EPP).  This does not effect security, since it was 
shown~\cite{SP00,GL01} that there exist classes of EPPs such that
applying the EPP and measuring out a key can be securely converted to protocols where a key
is first measured out and then we apply classical error correction and privacy amplification on the raw key to obtain
a secure one.
We now explain how to convert the above unfeasible protocol to a feasible one which can be performed via 
LOCC.

First, in step (2'), for the $n{-}m_x{-}m_z$ which are not used for testing, the twisting and untwisting
cancel and therefore, do not need to be performed.  
Also, the measurement of bit errors via  $\sigma_z\otimes\sigma_z$ on $AB$ commutes with the 
twisting and untwisting, and therefore, the twisting and untwisting cancel.
Similarly, in step (3'), the measurement commutes with the untwisting, and therefore this 
untwisting is also unnecessary.  
Finally, for step (2'), untwisting the state, estimating the expected number of phase errors, 
and retwisting is
equivalent to estimating the twisted phase error rate  
via the operator $\Sigma_x = U_{ABA'B'} (\sigma_x \ot \sigma_x
\ot I_{A'B'}) U_{ABA'B'}^\dagger$.  Mercifully, our only remaining task is to find 
a way to estimate this error rate via LOCC, rather than via direct measuring of the global 
operator $\Gamma_x$.

To do this, we will first decompose $\Sigma_x$ in terms of products of observables which can be locally measured
by Alice and Bob.  We then show that this estimation of the observable in terms of these product observables
is a good estimation.  As will be explained shortly, this involves
adapting the quantum deFinetti theorem\cite{Renner05} and a Chernoff-like bound.

We can always decompose any observable into product observables.  In particular:
\bea 
	\Sigma_x 
	& = & U_{ABA'B'} (\sigma_x \ot \sigma_x \ot I) U_{ABA'B'}^\dagger
\\
	& = & \sum_{j_a,j_b=1}^{t} s_{j_a j_b} O_{j_a AA'} \otimes O_{j_b BB'}
\label{eq:gx}
\eea
where $\{O_{j}\}_{j=1}^t$ is a basis (trace-orthonormal) for hermitian
operators acting on $AA'$ and $BB'$, and $t = d^2 d'$.
Alice and Bob can now estimate the average value of $\Sigma_x$ by dividing
the $m_z$ samples into $t^2$ groups, and then
estimating individually  $O_{j_a AA'}$ and $O_{j_b BB'}$ on the $i$th test
system.  They then multiply their results publicly, and finally sum these
products over $i=1,\cdots,m_z/t^2$ with the coefficients given by \eq{gx}. 

%
The outcome of this LOCC estimation procedure will result in giving some 
emperical value for the average of $\Sigma_x$, which we call $\SgmBempnon$.
We want to compare $\SgmBempnon$
to the emperical value $\SgmAAempnon$
obtained from estimating
$\Sigma_x$ via a direct global measurement (which is the measurement that is
performed in the unfeasible yet secure modified Lo-Chau protocol).  
If the two values are close,
then we have shown that the LOCC measurement is a good estimation of $\Sigma_x$

Indeed $\SgmBempnon$ will be close to $\SgmAAempnon$ if the
entire $m_z$ sample systems are in a joint tensor-power state $\untrust^{\otimes n}$, and if
the number of systems we test is large enough.  This follows from \eq{gx} and the fact that for tensor
power states, we may regard each measurement as an independent event.  We can
then use the Chernoff bound  
which states that a random sample of $k$ independent 
measurements of an operator $O$ 
on state $\rho^{\otimes n}$
will converge exponentially fast in $k$ to its average value ${\bar O}=\tr(O\rho)$.
More precisely, the probability that  $|O - {\bar O}|\geq \delta$ decays as $\sim e^{-Ck\delta^2}$
for $C$ a positive constant.
In this case we know that the estimate of each of the $t^2$ local measurements
will converge exponentially fast to $\tr(\untrust O_{j})$ as we increase the number of tested systems
$k=m_z/t^2$.
%

However, in our current problem, Alice and Bob share $\untrust$ which
is {\em not} a tensor-power state, and each measurement cannot be considered to
be an independent event.  Fortunately, there is a sense 
in which a random sampling
of $m_z$ systems is close to tensor-power.
First, permutation
symmetry can be imposed on the protocol (since we can choose a random sample in any order), 
and second, since the
estimation involves only a small portion ($m_z$) of the entire $n$
systems, the exponential quantum deFinetti theorem \cite{Renner05}
states that the measured (reduced) state is close to a mixture of
``almost-tensor-power-states''.  This is captured by Theorem 2 of the appendix. 
We can now apply a Chernoff-like bound to these almost-tensor-power-states.
%
The exact analysis involves many adaptations of the results in
\cite{Renner05} and is given in the appendix as Theorem 1.
The result has consequences well beyond the current considerations.  Essentially,
any realizations of an observable (i.e. a decomposition of the operator
in terms of others), is a good one, in the sense that performing one kind of measurement on $m$
out of $n$ systems via one
realization of the measurement, will yield average values which are well correlated with the values
obtained by performing another realization of the measurement on the remaining $n-m$ systems. 
This is captured in Theorem 3 of the appendix.  We can apply
this to the current case to show that the probability that 
$|\SgmBempnon-\SgmAAempnon | > \delta$
can be made small. This says that the estimated twisted phase errors 
through measuring a sample via LOCC is correlated with the result we would
obtain if we made an ideal measurement of twisted phase errors on the rest of system.
Thus in terms of security, 
the only difference between the modified protocol, and that of Lo-Chau, is that
instead of Equation \ref{eq:lcrate} governing the accuracy of the phase error estimate, we have 
through Theorem 3
%
%
\bea 
\label{eq:twistedlcrate}
	& & \Pr 
|\SgmBemp-\SgmAAemp | > \delta
\nonumber \\ 
  &\leq & 2 e^{-\smfrac{(n-m_z)(r+1)}{2n} + \smfrac{1}{2} d^4 d'^2 \ln (n-m_z)}
\nonumber \\ 
  &+& (t^2+1) 2^{-\left[\smfrac{\delta^2}{36 t^2 d^2 d'} 
                   - H(\smfrac{rt^2}{m_z})\right] \smfrac{m_z}{t^2} 
		   + d' d^2 \log(\smfrac{m_z}{2t^2}+1)}
\nonumber \\ 
  &+& d' d^2 e^{-\smfrac{m_z \delta^2}{72 d'^2 d^4 t^2}}
\eea
where the three expressions in the upper bound respectively come from
the exponential quantum deFinetti theorem, the Chernoff bound, and
random sampling theory.  Here, $d$ is the dimension of the maximally entangled state
and we can take $d=2$, $d'$ the dimension of each ancilla on $A'B'$, 
and $r$ is some natural number we will take to be $\geq d^4d'^2\ln{n}$.  The superscripts
for the emperical values of $\Sigma_x$ refer to $\SgmBempnon$ being measured using $m$
systems while $\SgmAAempnon$ is measured on the remaining $n-m$.

This then proves security of the entire scheme, since the only significant change
from the unfeasible modified protocol
is a different method for
 estimating phase errors.  The calculation of security 
in terms of composable security parameters for QKD~\cite{compos} is given in \cite{HHHLO06}. 

We now touch on several issues which arise.  The protocol we have given, as with all entanglement
based protocols, relies on keeping the quantum state $\untrust$ from decohering throughout the procedure,
and it is therefor not currently practical.  
However, it can be converted to a prepare-and-measure protocol
where Alice prepares a state, sends it down a channel (which might have zero quantum capacity), and
then Bob measures the state right away.  The conversion adapts well known techniques and
is contained in \cite{HHHLO06} along with an example.

Next, in the above protocol, we considered 
verification of tensor powers of private states with
dimension two on $A$  i.e. $\gamma_2^{\otimes n}$ under general attacks.  It is straightforward
to extend this to the verification of private states of any dimension, 
and states where the twisting is close to tensor power.
It is not clear whether one can extend this to private states which are not tensor power
such as a single $\gamma_d$; as of yet we do not have a no-go theorem.  
This is quite different from verification of pure state entanglement
where the maximally entangled state of any dimension can be written as $|\Phi_2\>^{\otimes n}$
and we are thus always trying to verify something close to tensor power.  

Here, we considered a twisted version of the Lo-Chau scheme, but we could have just as
well considered twisted versions of other parameter estimation schemes.  
Indeed our protocol is not optimal in its use of resources and
it may be interesting to improve it.  Some potential avenues were noted in \cite{HHHLO06}.
A tomographic verification scheme was suggested originally in \cite{HHHO03}, and it may
be interesting to explore its efficiency.  It is simpler in the sense that one 
could just discard some states, and be left with almost-tensor-product states as in
\cite{Renner05}.

Finally, here we have 
demonstrated conceptually that quantum key distribution is not
equivalent to the ability to send quantum information.  However, we only know of a few
channels and set of states which have the property of offering security without allowing
quantum communication.  It would be very interesting to find other examples, and perhaps 
even more interesting to know whether there are any bound entangled states (and the corresponding
zero-capacity channels) which cannot produce a secure key.

\noindent {\bf Acknowledgments} We thank Daniel Gottesman and
Hoi-Kwong Lo for valuable discussions. We acknowledge 
support from EU grant QAP IST-015848 and IP SCALA 015714.  
JO also acknowledges the Royal Society and DL is supported by 
the CRC, CRC-CFI, ORF, CIAR, NSERC, MITACS, and ARO.  KH acknowledges
the support of the Foundation for Polish Science.
\newpage
\section*{Appendix}

In Section \ref{sec:almostiid} we present Theorem 1 on the extent to which
a permutationally invariant set of systems behaves like independently and identically 
distributed (IID) states for the purpose of parameter estimation.  This is an application of
the quantum de Finetti theorem and the generalized Chernoff bound.  
Section \ref{sec:otherresults} presents some results which will be used in Section
\ref{sec:estimation}.  It is in this latter section where
the key Theorem 3 is presented in Subsection \ref{ss:mainthm}.  It relates the distance between
direct measurements and indirect measurements of an observable.
%
%

\section{LOCC estimation of the expectation of an IID observable}
\label{sec:almostiid}
\subsection{Finite quantum de Finetti theorem and generalized Chernoff bound}

We say that a state $\rho_n$ on Hilbert space $\hcal^{\ot n}$
satisfies the Chernoff bound with respect to a state $\sigma$ on
$\hcal$ and a measurement $\mcal$ on $\hcal$ if (with high
probability) the {\em frequency distribution} obtained by measuring
$\mcal^{\ot n}$ on $\rho_n$ is close to that of measuring $\mcal$ on
$\sigma$.  For example, 
$\rho_n = \sigma^{\ot n}$.
However many other states satisfy the same property.  An important
class is called {\em almost power states}, which are formulated and
studied in \cite{Renner05}.  We adapt results in \cite{Renner05} for our
own purpose in the following.

\begin{theorem}{\bf (Finite quantum de Finetti theorem plus Chernoff bound)}
\label{thm:fincher}
Consider any permutationally invariant (possibly mixed) state
$\rho_{n+k}$ on Hilbert space $\hcal^{\ot (n+k)}$. Let $\rho_n=\tr_k
\rho_{n+k}$ be the partial trace of $\rho_{n+k}$ over $k$
systems. Then there exists a probability measure $\mu$ on (possibly
mixed) states $\sigma$ acting on $\hcal$ and a family of states
$\rho^{(\sigma)}_{n,r}$ such that
\bee
\item The state $\rho_n$ is close to a mixture of the states
$\rho^{(\sigma)}_{n,r}$
\be
\label{ineqfin}
\left\| \rho_n - \int \rho^{(\sigma)}_{n,r} \; 
{d}\mu(\sigma) \right\|_{\rm tr} \leq 2\, 
e^{-{k(r+1)\over 2(n+k)} + {1\over 2} \dim(\hcal)^2 \ln k }
\ee
\item The states $\rho^{(\sigma)}_{n,r}$ (called {\rm almost power states}) 
satisfy the Chernoff bound in the following sense
\ben
\label{ineqcher}
{\rm Pr} \left( \left \|
P_{\mcal}(\sigma) - Q_{\mcal}(\rho^{(\sigma)}_{n,r})\right \| > \delta 
\right) 
\nonumber \\ \leq 
2^{-n \, \left[ {\delta^2 \over 4} - h({r\over n}) \right] 
+ |W|\log ({\smfrac{n}{2} + 1})} =:
e(\delta)
%
%
\een
where $\mcal{=}\{M_w\}_{w\in W}$ is any measurement on $\hcal$,
$P_{\mcal}(\sigma) = \{{\rm Tr}(\sigma M_w)\}_w$, 
$Q_\mcal(\rho^{(\sigma)}_{n,r})$ is the frequency distribution obtained
from measuring $\mcal^{\ot n}$ on the state
$\rho^{(\sigma)}_{n,r}$, and $|W|$ is the size of the alphabet $W$.
\item Reduced density matrices of the states $\rho^{(\sigma)}_{n,r}$
(to $n' \leq n$ systems) satisfy the same Chernoff bound:
\ben
{\rm Pr}( \left \| 
P_{\mcal}(\sigma) - Q_{\mcal}(\rho^{(\sigma)}_{n,r,n'}) \right \| > \delta) 
\nonumber \\ \leq 
2^{-n' \, \left[ {\delta^2 \over 4} - h({r\over n'}) \right] 
+ |W|\log ({\smfrac{n'}{2} + 1})}
\een
where $r \leq n'/2$ and $\rho^{(\sigma)}_{n,r,n'}=\tr_{n-n'}
\rho^{(\sigma)}_{n,r}$ is the partial trace of $\rho^{(\sigma)}_{n,r}$
over $n-n'$ systems.
\eee
\end{theorem}

\noindent {\em Proof:} We first collect various facts, definitions,
and results from \cite{Renner05}.

\subsubsection{Facts and definitions} 

{\definition {\bf Almost power state: (Def.\ 4.1.4, in \cite{Renner05})}
Suppose $0 \leq r\leq n$.  
Let $\; \Sym(\hcal^{\ot n})$ denote the symmetric
subspace of pure states of Hilbert space $\hcal^{\ot n}$.  
Let $|\theta \> \in \hcal$ be an arbitrary pure state and consider: 
\bea
\vcal(\hcal^{\otimes n}, |\theta\>^{\ot n - r}) := 
\{\pi(|\theta\>^{\ot n-r}\ot |\psi_r\>): 
\pi \in S_n, \, 
\nonumber 
\\|\psi_r\> \in \hcal^{\ot r} \}
\nonumber 
\eea
where $S_n$ is the permutation group of the $n$ systems.
We define the {\it almost power states along $|\theta\>$} to be
the set of pure states in 
\be 
|\theta\>^{[\otimes,n,r]} 
:= \Sym(\hcal^{\ot n}) \cap 
{\rm span}(\vcal(\hcal^{\ot n}, |\theta\>^{\ot n - r}))
\ee
We denote the set of mixtures of {\it almost tensor power states}
along $|\theta\>$ as ${\rm conv}(|\theta\>^{[\otimes,n,r]})$.}

\noindent With the above definition, we shall prove the following lemma: 

{\lemma ~~ 
If $\varrho_{n} \in {\rm conv}(|\theta\>^{[\otimes,n,r]})$, then,
$\varrho_{n-m} \in {\rm conv}(|\theta\>^{[\otimes,n-m,r]})$ where
$\varrho_{n-m}={\rm Tr}_m(\varrho_{n})$ is the reduced density matrix
after the partial trace over any $m$ out of the $n$ systems (by
symmetry, without loss of generality, we take the first $m$ systems).
\label{PSym}
}

{\it Proof .-} 

Since membership in ${\rm conv}(|\theta\>^{[\otimes,n-m,r]})$ is
preserved under mixing, it suffices to prove the lemma for pure
$\varrho_{n} = |\Psi\>\<\Psi|$, with $|\Psi\> \in 
|\theta\>^{[\otimes,n,r]}$.

We can pick an ensemble realizing $\varrho_{n-m}$ of our choice, and
prove the lemma by showing that any element $|\Psi_{n-m}\>$ in that
ensemble belongs to $|\theta\>^{[\otimes,n-m,r]}$.
Our ensemble is obtained by an explicit partial trace of $|\Psi_{n}\>$
over the first $m$ subsystems along the computational basis.  An element 
is given by
\be
|\Psi_{n-m}\rangle=\<i_{1}|...\<i_{m}| \otimes I_{n-m}|\Psi_{n}\rangle.
\label{AnsambleVector}
\ee
Now, we note two facts: 

(i) $|\Psi_{n-m}\> \in \Sym(\hcal^{\otimes(n-m)})$ -- This is because
$|\Psi_{n}\rangle \in \Sym(\hcal^{\otimes(n)}) = {\rm
span}(|\phi\>^{\otimes n})$.

(ii) $|\Psi_{n-m}\> \in \vcal(\hcal^{\otimes n{-}m},
|\theta\rangle^{\otimes (n-m-r)})$ -- This is because $|\Psi_{n}\> \in
\vcal(\hcal^{\otimes n}, |\theta\>^{\ot n - r})$, and expressing
$|\Psi_{n}\>$ in terms of the spanning vectors of
$\vcal(\hcal^{\otimes n}, |\theta\>^{\ot n - r})$ and putting it into
Eq.~(\ref{AnsambleVector}), we have
\be
|\Psi_{n-m}\rangle = \sum_{\Psi_{r},\pi} \alpha_{\Psi_{\!r},\pi} 
 \<i_{1}| \cdots \<i_{m}| \otimes I_{n-m} \; {\pi}  \, 
(|\theta\rangle^{\otimes n-r} \otimes
 |\Psi_{r}\rangle).
\nonumber
\ee
Elementary analysis shows that any term of the above sum is, up to
permutation, of the form
 $(\langle i_{1}|\theta \rangle) \cdots (\langle i_{p}|\theta \rangle) 
|\theta\>^{\otimes n-r-p} \otimes
[\langle i_{p+1}| \cdots \langle i_{m}| \otimes I_{r-(m-p)} \; \pi'  
(|\Psi_{r}\rangle)] $ 
where $0 \leq p \leq m$, and ``absorbing'' $m-p$ copies of $\theta$ to
the last part of the vector, we get
$|\theta\>^{\otimes n-(r+m)} \otimes |\Psi''_{r}\>$.  
Thus, $|\Psi_{m-n}\>$ is a sum of terms of the form
$\pi(|\theta\>^{\ot n-(r+m)} \otimes |\Psi''_{r}\>)$, and belongs to
$\vcal(\hcal^{\otimes (n-m)}, |\theta\>^{\ot n - (r+m)})$.  
This proves the second fact, and also the lemma.  $\square$

Property of a mixture of almost tensor power states behaves
approximately like a mixture of tensor power states, so that the
generalized version of Chernoff bound holds.  

{\lemma {\bf (Theorem 4.5.2 of \cite{Renner05})} Let ${\cal M} =
\{M_w\}_{w\in {\cal W}}$ be a POVM on ${\hcal}$, let $0\leq r \leq
\frac{n}{2}$.  Moreover let $|\theta\> \in \hcal$ and let
$|\Psi_{n}\>$ be a vector from $|\theta\>^{[\otimes,n,r]}$.  There
holds:
\ben
P(||P_{\mcal}(|\theta\rangle \langle \theta|) -
P_{\mcal}[|\Psi_{n}\rangle \langle \Psi_{n}|]|| > \delta) \nonumber
\\\leq 2^{-n \, \left[ {\delta^2 \over 4} - h({r\over n}) \right] 
+ |W|\log ({\smfrac{n}{2} + 1})}
\nonumber =: e(\delta)
\een
where $P_{\mcal}(|\theta\rangle\langle\theta|) = \{{\rm Tr} |\theta \rangle
\langle \theta|M_w\}_w$ and $P_{\mcal}[|\Psi_{n}\rangle\langle
\Psi_{n}|]$ is the frequency distribution of outcomes of ${\cal
M}^{\ot n}$ applied to $|\Psi_{n}\rangle \langle \Psi_{n}|$, and the 
probability is taken over those outcomes.  Note that we have used 
$e(\delta)$ instead of $\delta(e)$ in \cite{Renner05}.   
\label{GenChernoff}
}
 
Consider the general probability ${\rm Pr}( \| P_{\mcal}(\rho) -
P_{\mcal}[\varrho_{n}] \|<\delta)$ where $P_{\mcal}[\varrho_{n}]$ is a
frequency distribution of outcomes of ${\cal M}^{\ot n}$ applied to
$|\Psi_{n}\rangle \langle \Psi_{n}|$.  The distribution
$P_{\mcal}[\varrho_{n}]$, if treated as a functional of $\varrho_{n}$
on the space $\hcal^{\ot n}$, is {\it linear} in $\varrho_{n}$.
Following this we get immediately: 

\bec Lemma \ref{GenChernoff} holds when replacing the projector
$|\Psi_{n}\rangle \langle \Psi_{n}|$ (for $|\Psi_{n}\> \in
|\theta\>^{[\otimes,n,r]}$) by $\varrho_{n} \in {\rm
conv}(|\theta\>^{[\ot, n,r]})$.
\label{GenChernoffMixed}
\eec
 
Apart form the generalised Chernoff-type lemmas, we also need the
crucial exponential quantum finite deFinetti theorem:

\begin{theorem}[Theorem 4.3.2 of \cite{Renner05}]
For any pure state $|\psi_{n+k}\> \in \Sym(\hcal^{\ot n+k})$ and $0\leq
r\leq n$ there exists a measure $d\nu (|\theta\>)$ on $\hcal$ and for
each $|\theta\> \in \hcal$ a pure state $|\psi_n^{(\theta)}\> \in
|\theta\>^{[\ot, n,r]}$ such that
\ben
& & \left\| 
{\rm Tr}_k |\psi_{n+k}\>\<\psi_{n+k}| -  \int_{\hcal} 
|\psi_n^{(\theta)}\>\<\psi_n^{(\theta)}|  d\nu (|\theta\>)
\right\|_{\rm tr} 
\nonumber \\
& \leq & 2e^{-{k(r+1)\over 2(n+k)} + {1\over 2} \dim(\hcal)\ln k } 
\label{ineqfor_convhull}
\een
\label{th:for_convhull}
\end{theorem}
Finally, we need the fact that any permutationally invariant 
state has a symmetric purification.

\bel[Lemma 4.2.2 of \cite{Renner05}]
\label{lem:symmpurif}
Let $\rho_n$ be permutationally invariant state on $\hcal$. 
Then there exists purification of the state on 
$\Sym((\hcal\ot \hcal)^{\ot n})$
\eel

This concludes the list of facts and definitions needed for proving 
Theorem \ref{thm:fincher}.  

\subsubsection{Proof of Theorem \ref{thm:fincher}}

Consider an arbitrary permutationally invariant state $\varrho_{n+k}$ on
Hilbert space $\hcal^{\ot (n+k)}$.  \\
Step (1): According to Lemma
\ref{lem:symmpurif} there is a purification $|\psi_{n+k}\>$ that
belongs to $\Sym(\hcal'^{\ot n+k})$ where $\hcal'=\hcal \otimes
\tilde{\hcal}$ and dim$(\tilde{\hcal})=$dim$(\hcal)$. \\
Step (2): We apply to $\psi_{n+k}$ theorem \ref{th:for_convhull} with
the changes 
\bea
\hcal & \rightarrow & \hcal'=\hcal \otimes \tilde{\hcal}
\nonumber
\\
d & \rightarrow & d^2
\eea
Step (3): After application of theorem \ref{th:for_convhull} we perform
partial trace over $\tilde{\hcal}^{\ot n}$, the purifying spaces
introduced in (1).  We denote this partial trace by $\tilde{\rm Tr}$.
This partial trace induces from the measure on $\hcal'$ in step
(2) the new measure $\mu(\sigma)$ on the set of all mixed states $\sigma$
defined on $\hcal$.  (This is defined by probability ascribed by the
measure $\mu$ to the subset of $\hcal'$ equal to the
equivalence class of all pure states $|\theta\>$ which satisfy
$\tilde{\rm Tr}(|\theta\rangle\langle \theta|)=\sigma$).  This partial
trace produces also the states $\almpower$ defined directly by
$\almpower \equiv \tilde{\rm Tr}(|\psi_n^{(\theta)}\rangle \langle
\psi_n^{(\theta)}|)$ where the existence of the pure states
$|\psi_n^{(\theta)}\>$ is guaranteed by theorem \ref{th:for_convhull}.
Finally we note that partial trace does not increase the trace
distance between two quantum states, so applying partial trace to the
LHS of (\ref{ineqfor_convhull}) and using the notation described above
we get immediately the inequality (\ref{ineqfin}).  This proves the
first item of Theorem (\ref{thm:fincher}).

To prove the second item of Theorem (\ref{thm:fincher}), remember from
the above that $\almpower \equiv \tilde{\rm
Tr}(|\psi_n^{(\theta)}\rangle \langle \psi_n^{(\theta)}|)$.  Since
$|\psi_n^{(\theta)}\rangle$ is an almost power pure state, lemma
\ref{GenChernoff} applies.  Further, it holds for all POVM-s on
$\hcal'=\hcal\ot\tilde{\hcal}$, and in particular for incomplete
POVM-s acting only on $\hcal$ but not on $\tilde{\hcal}$.  Thus, the
conclusion of lemma \ref{GenChernoff} holds with the change: $\mcal
\rightarrow \mcal \otimes I$, which gives item (2).  

Finally, to prove item 3 of theorem \ref{thm:fincher}, note that the
reduced density matrices $\varrho_{n,r,n'}^{\sigma}$ of interest can be
obtained from the pure state $|\psi_n^{(\theta)}\>$ above by tracing
(i) first over $n-n'$ subsystems corresponding to $\hcal'$, producing a
state on $\hcal'^{\ot n'}$, and
(ii) then over $n'$ subsystems corresponding to $\tilde{\hcal}$.

Then lemma \ref{PSym} guarantees that the first partial trace produces
a mixed state $\varrho_{n'}$ in ${\rm
conv}(|\theta\>^{[\ot,n',n'-r]})$ (with underlying space $\hcal'$.
Applying corollary \ref{GenChernoffMixed} to $\varrho_{n'}$ with $n'$
instead of $n$, it suffices to consider a pure state in
$|\theta\>^{[\ot,n',n'-r]}$.  Finally, lemma \ref{GenChernoff} can be
applied to this pure state with $\mcal \rightarrow \mcal \otimes I$
which concludes item 3 (with the assumption $0\leq r\leq
\frac{n'}{2}$).  $\square$

\subsection{Two other useful results} 
\label{sec:otherresults}
\subsubsection{Classical random sampling} 

In addition to the fact and definitions above and Theorem
\ref{thm:fincher}, we will need the following result on classical
random sampling (or equivalently symmetric probability distribution). 

\bep{\bf(Classical sampling theory) Lemma A.4 from \cite{RK04b}.}
\label{prop:srodka}
Let $Z$ be an $n$-tuple and $Z'$ a $k$-tuple of random variables over
 set $\zcal$, with symmetric joint probability $P$. Let $Q_{z'}$ be a
 frequency distribution of a fixed sequence $z'$ and $Q_{(z,z')}$ be
 frequency distribution of a sequence $(z,z')$. Then for every $\ep
 \geq 0$ we have \be P(||Q_{(z,z')} - Q_{z'}||\geq \ep )\leq |\zcal|
 e^{-{k\ep^2/8|\zcal|}}.  \ee \eep The result says that frequency
 obtained from a small sample is close to frequency distribution
 obtained from the whole system.

\subsubsection{From probabilities to averages}

{\lemma Consider an observable $L$ on Hilbert space $\hcal$,
$\dim\hcal=d$. Let $L=\sum_{i=1}^{t} s_i L_i$, where $L_i$ satisfy
$\tr L_i L_j^\dagger=\delta_{ij}$.  Let eigenvalues of $L_i$ be
denoted by $\lambda_l^{(i)}$.  Consider arbitrary state $\rho$, and
let $P^{(i)}=\{p_l^{(i)}\}$ be the probability distribution on $l$
(which eigenvalue) induced by measuring $L_i$ on $\rho$.  Let
$Q^{(i)}=\{q_l^{(i)}\}$ be an arbitrary family of distributions on
eigenvalues of $L_i$.  We then have
\bea
\nonumber
& & |\<L\>_{\rho} - \sum_{i} s_i \sum_l \lambda_l^{(i)}q^{(i)}_l | \\
& \leq & 
\sqrt{t}||L||_{HS} \max_i ||P^{(i)}-Q^{(i)}||,
\eea
where $\| \cdot \|_{HS}$ is the Hilbert-Schmidt norm, $\| \cdot
\|_\infty$ is the operator norm, and $\| \cdot \|$ is the trace norm.
\label{closeaverages}
}\\
{\it Proof}
\ben
&& \left|
\<L\>_{\rho} - \sum_{i} s_i \sum_l \lambda_l^{(i)} q^{(i)}_l \right|
\nonumber
\\
&=&
\left|
\sum_i s_i \sum_l \lambda_l^{(i)}(p_l^{(i)} {-} q_l^{(i)}) \right| 
\nonumber
\\
&\leq& 
\sum_i |s_i| \; (\max_l |\lambda_l^{(i)}|) \, \| P^{(i)}-Q^{(i)} \| 
\nonumber
\\
&= & 
\sum_i s_i \| L_i \|_{\infty} \, \| P^{(i)}-Q^{(i)} \| 
\nonumber
\\
& \leq & 
( \max_j \| P^{(j)}-Q^{(j)} \|) \sum_i s_i \; \| L_i \|_{\infty} 
\een
Since $||L_i||_{\infty} =1$, using convexity of $x^2$ we obtain 
\be
\sum_{i=1}^t \, s_i \, \| L_i \|_{\infty}
=\sum_i s_i \leq \sqrt{t} \sqrt{\sum_i s_i^2}=\sqrt{t} \, \|L\|_{HS}
\ee
which ends the proof. $\square$ 
%

\subsection{Estimation - detailed description}
\label{sec:estimation}
We consider $2m+n$ systems with Hilbert space $\hcal^{\ot {(2m +n)}}$,
$\dim \hcal = d$ in a permutationally invariant state $\varrho_{2m+n}$.
%
%
Suppose the ultimate goal is to obtain the ``frequency mean-value'' of
some single-system observable $\Sigma$ on $n+m$ systems. 
In other words, we want to measure
$\frac{1}{N}\sum_{j=1}^{N}\Sigma^{(j)}$ where $\Sigma^{(j)}=I \otimes
I \otimes \cdots \otimes \Sigma \otimes \cdots \otimes I$ on the $N$
subsystems for $N=n+m$.
 
Because of experimental limitations (here, it is the LOCC constraints
on Alice and Bob), they are restricted to measuring product operators of the
form $L = L_A \otimes L_B$ by independently finding the eigenvalues of
$L_A$ and $L_B$ (i.e., making the measurements $L_A \otimes I$ and $I
\otimes L_B$), discussing over classical channels and multiplying
their outcomes together.  
Now, to measure $\Sigma$, one can first rewrite it in terms of 
product operators $L_i$: 
\begin{equation}
\Sigma=\sum_{i=1}^{t} s_{i} L_{i}
\label{eq:local_decomp}
\end{equation}
where we have chosen $\{L_i\}$ to be hermitian and trace orthonormal, so 
that $s_i$ are real.  The $L_i$-s are ``{\em intermediate observables}.''
We will describe an inference scheme that (1) involves only the
estimation of the ``frequency mean-value'' of $\Sigma$ on a small
number ($m$) of subsystems, and (2) the measurement of $\Sigma$ is
done indirectly via measurements of the $L_i$'s. 

The analysis will start with special assumption about the $2m$-element
sample, $m$ of which are used for indirect estimation.  The
assumptions are relaxed on that sample.  After that properties of the
other $m+n$ subsystems are inferred.

\subsubsection{Analysis of the $2m$ sample 
in an ``almost power state along $\sigma$'':$\Romr$}
 
Suppose the first $2m$ subsystems are in a joint state $\Romr$, with
$r\leq \frac{1}{2} \times 2m$.  We expect the state $\Romr$ to play a
role similar to the state $\sigma^{\ot 2m}$.
Define the theoretical direct average 
\begin{equation}
\Sgm={\rm Tr}(\Sigma \sigma)=\sum_{i}s_{i}\langle L_{i}\rangle_{\sigma}
\label{eq:ss}
\end{equation}
We will show that the empirical average, either obtained directly or
indirectly, will be close to the above.  

For the indirect measurement, divide the first $m$ subsystems into
$t$ groups. Each group has $m'=m/t$ subsystems.
Alice and Bob take the $i$th group ($i=1,\cdots,t$) and measure $L_i$ on each
site as described above (the measurement is ${\cal L}_{i}$).
In other words, the measurement $M^{\rm indirect} = \otimes_{i=1}^{t}(
{\cal L}_{i}^{\otimes m'})$ is applied to the first $m$ subsystems of
the entire $2m+n$ subsystems.
The reduction of the state $\Romr$ to the first $m$ subsystems induces
probability distribution $\pcal$ on the outcomes of $M^{\rm
indirect}$.

Since we expect $\Romr$ and $\sigma^{\ot m}$ to behave similarly,
consider the probability distribution on alphabet $\acal_i$ of
observable $L_{i}=\sum_{l}\lambda_{l}^{i} \overline{P}_{l}^{(i)}$
induced by the state $\sigma$ as follows:
\begin{equation}
P_{i}=\{ {\rm Tr}(\sigma \overline{P}_{l}) \}_l
\end{equation} 
An execution of the measurement ${\cal L}_{i}^{\otimes m'}$ gives a
particular outcome $(l_{1},...,l_{m'})$ and induces frequency
distribution $Q_{i}$ on alphabet $\acal_i$ of the observable $L_{i}$.

Then, the empirical frequency distributions $Q_i$ is close to the
``theoretical'' distribution $P_i$: 
\begin{fact}
\label{fact:distributions}
\be
{\cal P}(\| P_{i}-Q_{i} \| \geq \delta)\leq e(\delta,m',r,d),
\ee
where $d$ is the dimension of the single site Hilbert space, and 
\begin{equation}
 e(\delta,n,r,|Z|) \; {:}{=} \; 2^{-(\frac{\delta^{2}}{4}-H(\frac{r}{n}))n +
 |Z| \log(\frac{n}{2}+1)}
\end{equation}
\end{fact}

{\it Proof -} Follows immediately from the third item of Theorem
\ref{thm:fincher}.  Note that we use item (3) not (2) since we perform
the measurement only on {\it part} of the state $\Romr$.

{\it Remark -} Note also that $P_{i}$ is constant while $Q_{i}$
is a random variable.

Now, we define the theoretical average values for the intermediate 
observables $L_i$'s:
\be
\Li = {\rm Tr}(L_{i}\sigma)
\ee
and the empirical average 
\be
\Liemp=\sum_{l}\lambda_{l}^{(i)}Q_{i}(l)
\ee
where $Q_i(l)$ denotes value of $Q_i$ on specific event $l$ from
alphabet $\acal_i$. (Again, $ \langle L_{i} \rangle_{\sigma}$ is
constant while $\langle L_{i} \rangle_{emp}$ is a random variable
depending on the particular outcomes of measurement - recall that
$L_{i}=\sum_{i}\lambda_{l}^{(i)}\bar{P}_{l}^{(i)}$ ).
And again, recall that we the {\it empirical} value of $\Sigma$
obtained {\it indirectly}, {\it via} empirical distributions of
the $L_{i}$.
\begin{equation}
\SgmBemp = \sum_{i}s_{i}\langle L_{i}\rangle_{emp} \,.
\end{equation}
We now show that the indirect empirical average is close to the 
direct theoretical average in \eq{ss}.  
First applying the union bound to Fact \ref{fact:distributions}, we get
\begin{equation}
{\cal P}(\cup_{i=1,...,t}\{||P_{i} -Q_{i}||>\delta  \})\leq
t\cdot e(\delta,m',r,d)
\end{equation} 
Then using Lemma \ref{closeaverages}  we obtain 
that $P(|\sum_{i}^{t}s_{i}\Li -\sum_{i}^{t}s_{i}\Liemp|>\delta)
\leq t \cdot e(\frac{\delta}{||\Sigma||_{HS}\sqrt{t}},m',r,d)$
which is just
\be
\label{Ineq1}
{\cal P}(|\Sgm-\SgmBemp|>\delta)\leq t \cdot
e(\frac{\delta}{||\Sigma||_{HS}\sqrt{t}},m',r,d).
\ee

After considering the indirect measurements, suppose that someone
measures directly $M^{\rm direct} = \sum_{j=m+1}^{2m} \Sigma^{(j)}$ on
the second group of $m$ subsystems.
The empirical average outcome is given by
\begin{equation}
\SgmAemp=\sum_{x}\gamma_{x} {Q}(x)
\end{equation}
where $Q$ is the frequency distribution on the alphabet of $\Sigma$
(similarly as $Q_i$ is the frequency distribution of alphabet
$\acal_i$ of $\lcal_i$), and $\gamma_{x}$ are some real numbers.
In a way similar to the indirect case (but much easier here) we show 
that the empirical direct average is close to \eq{ss}: 
\be
\label{Ineq2}
{\cal P}(|\Sgm - \SgmAemp|>\delta)\leq
e(\frac{\delta}{||\Sigma||_{HS}},m,r,d).  
\ee

{From} the inequalities (\ref{Ineq1}), (\ref{Ineq2}) we obtain
\bel
\label{lem:SigmaDistance}
For the measurements on the state $\Romr$ 
considered above we have:
\bea
& & {\cal P}(|\SgmAemp-\SgmBemp|>2 \delta) \nonumber 
\\ & \leq & t \cdot e(\frac{\delta}
{||\Sigma||_{HS}\sqrt{t}},m',r,d) +
e(\frac{\delta}{||\Sigma||_{HS}},m,r,d)
\nonumber \\ 
& \leq & (t+1)
e(\frac{\delta}{||\Sigma||_{HS}\sqrt{t}},m',r,d)  
\eea
\eel

{\it Proof .-} Here triangle inequality and union bound to
inequalities (\ref{Ineq1}), (\ref{Ineq2}) suffices together with the
properties of $e(\delta,n,r,d)$.


\subsubsection{Passing from $\Romr$-s to their integrals and then to 
a close-by state}

Note that both integration and the measurement of a state to produce
the classical distribution of the outcomes are both linear, completely
positive, and trace-preserving maps.  
Thus, Lemma \ref{lem:SigmaDistance} still holds under the replacement
$\Romr \rightarrow \int \Romr d\mu(\sigma)$.
Furthermore, if 
\be
||\varrho_{2m} - \int \Romr d \mu(\sigma)|| \leq \epsilon
\ee
and since the trace distance is nonincreasing under the measurement (a
TCP map), the output distribution is different by no more than $\epsilon$.  
In this way we have proven 
 \bel
 \label{lem:StateDistance}
 For a state $\varrho_{2m}$ of $2m$ systems 
 satisfying $||\varrho_{2m} - \int \Romr d \mu (\sigma)||\leq \epsilon $
 we have 
 \bea
 & & {\cal P}'(|\SgmAemp-\SgmBemp|>2\delta) 
\nonumber
\\  & \leq & 
 (t+1)e(\frac{\delta}{||\Sigma||_{HS}\sqrt{t}},m',r,d) +\epsilon.
 \eea
 where ${\cal P}'$ is the probability distribution on outcomes of
 measurement ${\cal L}_{1}^{\otimes m'} \otimes ...\otimes {\cal
 L}_{t}^{\otimes m'} \otimes {\cal M}^{\otimes m}$ induced by the state
 $\varrho_{2m}$.
\eel

\subsubsection{Inferring direct average 
on $n+m$ samples of general state $\varrho_{2m+n}$ from 
indirect measurements on $m$ samples}
\label{ss:mainthm}
Now we pass to the general permutationally invariant 
state $\varrho_{2m+n}$.  
We want to relate the distance between  $\SgmBemp$,
the indirect estimation of $\Sigma$ obtained via LOCC measurements 
$\{{\cal L}_i\}$ on $m$ of the systems,
and the direct estimation $\SgmAAemp$ of $\Sigma$, we would obtain via
the direct measurement $\cal M$ on the other $n+m$ systems.
We have the following: 
\bet
\label{thm:main}
Consider permutationally invariant state $\varrho_{2m+n}$ on
$\hcal^{\ot 2m+n}$ and $\dim \hcal = d$. On this state we perform the
measurement ${\cal L}_{1}^{\otimes m'} \otimes \cdots \otimes {\cal
L}_{t}^{\otimes m'} \otimes {\cal M}^{\otimes m+n}$ which induces the
probability measure ${\cal P}''$. (Note that $\pcal'$ from Lemma \ref{lem:StateDistance} 
is simply the marginal of $\pcal''$.)  Then we have
\be
{\cal P}''(|\SgmBemp-\SgmAAemp)|> 3 \delta) \leq e_{1} + e_{2} +e_{3}
\ee
where 
\be
e_{1}=2e^{-{n(r+1)\over 2(2m+n)} + {1\over 2} d^{\bf 2} \ln n },
\ee
\be
e_{2}=(t+1)
2^{-(\frac{\delta^{2}}{4t||\Sigma||_{HS}^{2}}-H(\frac{r}{m'})) m'  
+ d \log(\frac{m'}{2}+1)}
\ee
and
\be
e_{3}=d e^{-\frac{m\delta^{2}}{ {\bf 8}d||\Sigma||_{HS}^{2}}}.
\ee
where $||\cdot||_{HS}$ is the Hilbert-Schmidt norm. 
\label{th:box2}
\eet
%


{\it Proof -} The parameters $e_{1}$,$e_{2}$, $e_{3}$ come 
from the generalised quantum de Finetti theorem, Chernoff bound 
and sampling proposition respectively. 

To start with the proof note that from Theorem \ref{thm:fincher}, 
item 1 we get that for 
$\varrho_{2m}={\rm Tr}_{n}\varrho_{2m+n}$ we have
$||\varrho_{2m}-\int \Romr d \mu(\sigma)|| \leq \epsilon $
 with $\epsilon=e_{1}$.
Applying then Lemma  \ref{lem:StateDistance} we get that 
\be
{\cal P}'(|\SgmAemp-\SgmBemp|>2 \delta) \leq e_{1} + e_{2}
\label{e12}
\ee

Now we need to connect $\SgmBemp$ with $\SgmAAemp$.  For this we need
sampling Proposition \ref{prop:srodka} ${\cal P}''(||Q_{\Sigma}^{m} -
Q_{\Sigma}^{m+n}|| > \delta )\leq d e^{-{m \delta^2/8 d}}$ where
$Q_{\sigma}^{m}$ is the frequency distribution on outputs of $M$ induced
by the state $\rho_m$ (partial trace of $\rho_{2m+n}$ over $m+n$
systems and $Q_{\sigma}^{m+n}$ is frequency distribution induced on
outcomes of $M$ by state $\rho_{m+n}$ (partial trace of $\rho_{2m+n}$
over $m$ systems) and $d$ is the dimension of elementary Hilbert space
${\cal H}$ (thus $\varrho_{2m}$ is defined on ${\cal H}^{\otimes 2m}$.
Using Lemma \ref{closeaverages} we go to the averages \be {\cal
P}''(|\SgmAemp-\SgmAAemp|>3\delta) \leq e_{3}.
\label{e3}
\ee
Applying the union bound to Eqs.\ (\ref{e12}) and (\ref{e3})
we get finally the statement of the theorem.


\bibliographystyle{Science}


\end{document}